\documentclass[usenatbib]{mn2e}
\usepackage{times}
\usepackage{graphicx}

\def\mnras{MNRAS}
\def\apj{ApJ}
\def\apjs{ApJ Supplement Series}
\def\apjl{ApJ Letters}
\def\physrep{Phys Reps}
\def\aj{A J}
\def\aap{A\&A}

\def\x{{\bf x}}

\def\g{\gamma}
\def\th{\Theta}

\pubyear{2008}

\author[Bagla, Yadav and Seshadri]
{J. S. Bagla$^1$, Jaswant Yadav$^2$ and T. R. Seshadri$^3$ \\
  $^1$ Harish-Chandra Research Institute,  Chhatnag Road, Jhusi,
  Allahabad 211019, India. \\
  $^2$, $^3$ Department of Physics and Astrophysics, University of Delhi,
  Delhi 110007, India \\
  E-mail: $^1$ jasjeet@hri.res.in, $^2$ jaswant@physics.du.ac.in, $^3$
  trs@physics.du.ac.in}

\title[Fractal Dimensions of a Weakly Clustered Distribution]
{Fractal Dimensions of a Weakly Clustered Distribution and the Scale of
  Homogeneity}

\label{firstpage}

\def\LaTeX{L\kern-.36em\raise.3ex\hbox{a}\kern-.15em
    T\kern-.1667em\lower.7ex\hbox{E}\kern-.125emX}

\pagerange{\pageref{firstpage}--\pageref{lastpage}}

\begin{document}

\maketitle


\begin{abstract}
Homogeneity and isotropy of the universe at {\sl sufficiently} large scales is
a fundamental premise on which modern cosmology is based.
Fractal dimensions of matter distribution is a parameter that can be used to
test the hypothesis of homogeneity.
In this method, galaxies are used as tracers of the distribution of matter and
samples derived from various galaxy redshift surveys have been used to
determine the scale of homogeneity in the Universe.
Ideally, for homogeneity, the distribution should be a mono-fractal with the
fractal dimension equal to the ambient dimension.
While this ideal definition is true for infinitely large point sets, this may
not be realised as in practice, we have only a finite point set.
The correct benchmark for realistic data sets is a homogeneous distribution of
a finite number of points and this should be used in place of the
mathematically defined fractal dimension for infinite number of points $(D)$
as a requirement for approach towards homogeneity.
We derive the expected fractal dimension for a homogeneous distribution of
a finite number of points.
We show that for sufficiently large data sets the expected fractal
dimension approaches $D$ in absence of clustering.
It is also important to take the weak, but non-zero amplitude of clustering at
very large scales into account.
In this paper we also compute the expected fractal dimension for a finite
point set that is weakly clustered.
Clustering introduces departures in the Fractal dimensions from $D$ and in
most situations the departures are small if the amplitude of clustering is
small.
Features in the two point correlation function, like those introduced by
Baryon Acoustic Oscillations (BAO) can lead to non-trivial variations in the
Fractal dimensions where the amplitude of clustering and deviations from $D$
are no longer related in a monotonic manner.
We show that in the concordance model, the fractal dimension makes a rapid
transition to values close to $3$ at scales between $40$ and $100$~Mpc.
\end {abstract}




\section{Introduction}

We expect the Universe to be homogeneous and isotropic on the largest scales.
Indeed, one of the fundamental postulates in cosmology is that the Universe is
spatially homogeneous and isotropic.
It is this postulate, generally known as the cosmological principle
(CP)\citep{1917SPAW.......142E}, that allows us to approximate the description
of space-time by a Friedman-Robertson-Walker-Lemaitre (FLRW) metric.
The standard approach to cosmology assumes that the universe can be modelled
as a perturbed FLRW universe.
The large scale structures (LSS) in the universe are believed to have been
formed due to the collapse of small inhomogeneities present in the early
Universe
\citep{1980lssu.book.....P,1999coph.book.....P,2002tagc.book.....P,2002PhR...367....1B}.
Thus it is of paramount importance to test whether the observed distribution
of galaxies approaches a homogeneous distribution at large scales.

The primary aim of galaxy surveys
\citep{2001MNRAS.328.1039C,2000AJ....120.1579Y,1996ApJ...470..172S} is to
determine the distribution of matter in our Universe.
Redshift surveys of galaxies have revealed that the universe consists of a
hierarchy of structures starting from groups and clusters of galaxies to
super clusters and interconnected network of filaments spread across the
observed Universe
\citep{2007arXiv0708.1441V,2000PhRvL..85.5515C,1986ApJ...302L...1D,2002AJ....123...20K}.

Fractal dimensions can be used as an indicator to test whether or not the
distribution of galaxies approaches homogeneity.
One of the reasons that make the Fractal dimensions an attractive option is
that one does not require the assumption of an average density
\citep{1982fgn..book.....M,2002sgd..book.....M}.
Ideally one would like to work with volume limited samples in order to avoid
corrections due to a varying selection function.
Redshift surveys of galaxies can be used to construct such sub-samples from
the full magnitude limited sample but this typically leads to a sub-sample
that has a much smaller number of galaxies as compared to the full sample.
This limitation was found to be too restrictive for the earliest surveys and
corrections for the varying selection function were attempted in order to
determine the scale of homogeneity: for example see
\citet{1999A&A...351..405B}.
With the large surveys available today, this limitation is no longer very
serious.
Fractal dimensions are computed for the given sample or sub-sample and the
scale beyond which the fractal dimension is close to the physical dimension of
the sample is identified as the scale of homogeneity.
We expect that at scales larger than the scale of homogeneity, any fluctuation
in density are small enough to be ignored.
Thus at larger scales, CP can be assumed to be valid and it is at these scales
that the FLRW metric is a correct description of the Universe.

Fractal Dimension is defined in the mathematically rigorous way only for an
infinite set of points.
Given that the observational samples are finite, there is a need to understand
the relation between the fractal dimension and the physical dimension for such
samples.
In this paper, we compute the expected fractal dimension for a finite
distribution of points (see e.g.,
\citep{1993PhRvE..47.3879B,1994PhRvE..49.4907B}).
The early work on these effects has focused on small scales where the
amplitude of clustering is large.
In this work, we calculate fractal distribution for a uniform distribution, as
well as for a weakly clustered distribution of a finite number of points.
This is of interest at larger scale where fractal dimensions are used as a
tool to find the scale of homogeneity.

Catalogues of different extra-galactic objects have been studied using various
statistical methods.
One of the important tools in this direction has been the use of two point
correlation function $\xi(r)$ \citep{1980lssu.book.....P} and its Fourier
transform the power spectrum $P(k)$.
We have precise estimates of $\xi(r)$
\citep{2007MNRAS.378.1196K,2006astro.ph.12400R,2007A&A...472...29D}  and the
power spectrum  $P(k)$ \citep{2005MNRAS.362..505C,2007ApJ...657..645P} from
different galaxy surveys.
Different measurements appear to be consistent with one another once
differences in selection function are accounted for
\citep{2006astro.ph.11178C} (but also see \citet{2007arXiv0708.1517S}).
On small scales the two point correlation function is  found to be well
described by  the form
\begin{equation}
  \xi(r)=\left(\frac{r_0}{r}\right)^{\g}
\label{eq:1}
\end{equation}
where  $\g =1.75\pm 0.03 $ and $ r_0=6.1\pm 0.2$~$h^{-1}$Mpc for the SDSS
(\citet{2002ApJ...571..172Z}) and $\g =1.67\pm 0.03 $ and $ r_0=5.05\pm
0.26$~$h^{-1}$Mpc for the 2dFGRS (\citet{2003MNRAS.346...78H}).
Recent galaxy surveys have reassured us that the power law behaviour for
$\xi(r)$ does not extend to arbitrary large scales.
The breakdown  of this behaviour occurs at  $r> 16 h^{-1} {\rm Mpc} $ for SDSS
and at $r> 20 h^{-1} {\rm Mpc} $ for 2dFGRS,
which is consistent with the distribution of
galaxies being homogeneous at large scales.
A note of caution here is that though the $\xi(r)$ determined from
redshift surveys is consistent with the universe being homogeneous at
large scales in that $\left|\xi(r)\right| \ll 1$ at large $r$, it does not
actually imply that the universe is homogeneous.
This is because the two point correlation given by,
\begin{equation}
\xi(r)=<\delta(x+r)\delta(x)>
\label{eq:2}
\end{equation}
where
\begin{equation}
\delta(x) = \frac{\rho(x) - \bar\rho}{\bar\rho}
\label{eq:3}
\end {equation}
presupposes that galaxy distribution that we
are analysing is homogeneous on the large scales of our survey. This is
implicit in the fact that $\bar\rho$, which is assumed to be the spatial
average density of matter in the universe, is computed by averaging the
density from within the survey volume.
Of course, it may be possible to demonstrate that the survey is a fair sample
of the universe by showing that the values of $\bar\rho$ derived from
sub-samples of the survey are consistent with each other, or that the value of
$\bar\rho$ computed at different scales converges to a definite value at
scales much smaller than the size of the survey.
To verify and hence validate the cosmological principle, it is useful
to consider a statistical test which does not presuppose the premise being
tested.
In this paper we consider one such test, the ``multi-fractal analysis''
and apply it to distribution of particles in random as well as clustered
distributions.

Fractal dimension, which is generally a fractional number, is
characterised by the scaling exponent. In most physical situations,
we need to use a set with an invariant measure characterised by a
whole spectrum of scaling exponents, instead of a single number.
Such a system is called a multi-fractal and we need to do a
multi fractal analysis of a point set to study the system.

Various groups have used the concept of fractals to analyse catalogues of
extra-galactic objects.
See \citet{2005RvMP...76.1211J} for an excellent review of quantitative
measures used for describing distributions of points.
Based on the scale invariance of galaxy clustering,
\citet{1987PhyA..144..257P} suggested that the distribution of galaxies is a
fractal to arbitrarily large  scales.
In a later analysis of different samples of galaxies
\citet{1992PhR...213..311C} obtained results consistent with this argument.
On the other hand \citet{1995PhR...251...1} showed that the distribution is a
fractal only on small scales and on large scales there is a transition to
homogeneity.
If the distribution of galaxies is found to be a fractal then the average
number of galaxies in a volume of radius $r$ centred on a galaxy should scale
as $r^{d}$, where $d$ is the fractal dimension.
Hence the number density of neighbouring galaxies would go as $\rho = r^{d -
  D}$ in a $D$ dimensional distribution.
This, when calculated for higher values of $r$ will show a decrease from that
of lower scales.
This effect led \citet{1998PhR...293...61S} to believe that the value of
correlation length $r_0$ (eq.\ref{eq:1}) increases with the increase in size
of the sample.
However this interpretation is not supported by volume limited samples of
various galaxy redshift surveys \citep{1996ApJ...472..452B,
  2001ApJ...554L...5M}.
A number of authors
\citep{1998A&A...335..779C,1999MNRAS.310.1128H,2000ApJ...541..519B,1999ApJ...514L...1A,
2004AstL...30..444B,2006A&A...447..431V,2007A&A...465...23S} have
shown the distribution of galaxies to be a mono-fractal up to the
largest scales that they were able to analyse. On the other hand
homogeneity has been seen at large scale in other analyses
\citep{1997NewA....2..517G,1999A&A...351..405B,1999Sci...284..445M,2001A&A...370..358K,2000MNRAS.318L..51P,2005ApJ...624...54H,2005MNRAS.364..601Y}.
The best argument in favour of large scale homogeneity stems from
the near isotropy of radio sources or background radiation in
projection on the sky \citep{1999Natur.397..225W}.

The aim of this paper is to calculate the fractal dimension for a distribution
of finite number of points which are distributed homogeneously as well as for
those which are weakly clustered.
For this purpose we use the multi-fractal analysis to study the scaling
behaviour of uniform as well as weakly clustered distributions in turn finding
the relationship between the fractal dimension and the two point correlation
function.
We find deviations of fractal dimension $D_q$ from the $D$ arising
due to a finite number of points for a random distribution with uniform
density, these deviations arise due to discreteness.
In this case we can relate the deviation (of $D_q$ from $D$) to the number
density of points.
We further show that for a distribution of points with weak clustering, there
is an additional deviation of $D_q$ from $D$.
This deviation can be related to the two point correlation and the intuitive
relation between the amplitude of clustering and deviation from homogeneity is
given a quantitative expression.
We then apply the derived relation to cosmology and compute the expected
deviations in a model that fits most observations.

A brief outline of the paper is as follows. In \S{2} we describe the method of
analysis, \S{3} contains results and discussion with the conclusions in \S{4}.


\section{Fractal Dimensions}

Fractal dimension is the basic characterisation of any point
distribution.
There are many different methods that can be used to calculate the fractal
dimension.
Box counting dimension of fractal distribution is defined in terms of non
empty boxes $N(r)$ of radius $r$ required to cover the distribution. If
\begin{equation}
 N(r) \propto r^{D_b}
\label{eq:4}
\end{equation}
we define $D_b$ to be the box counting dimension.
One of the difficulties with such an analysis is that it does not depend on
the number of particles inside the boxes and rather depends only on the number
of boxes.
As such it provides limited information about the degree of clumpiness of the
distribution and is a purely geometrical measure.
To get more detailed information on clustering of the distribution we use the
concept of correlation dimension.
Instead of using the formal definition of correlation dimension, which demands
that the number of points in the distribution approach infinity, we choose a
working definition which can be applied to a distribution of a finite number
of points.
Calculation of the correlation dimension requires the introduction of
correlation integral given by
\begin{equation}
C_2(r)=\frac{1}{NM}\sum_{i=1}^{M}n_i(r)
\label{eq:5}
\end{equation}
Here we assume that we have $N$ points in the distribution and we have $M$
cells centred on a fraction of these points.
In general the number of points and cells are different as one cannot use
points near the edge of the sample where a sphere of radius $r$ is not
completely inside the sample.
$n_i(r)$ denotes the number of particles within a distance $r$ from a
particle at the point $i$:
\begin{equation}
n_i(r)=\sum_{j=1}^{N}\th(r-\mid \x_i-\x_j \mid)
\label{eq:6}
\end{equation}
where  $\th(x)$ is the  Heaviside function.

For the purpose of the analysis in this paper, it will be useful to define
$C_2$ in terms of the probability of finding particles in a sphere of radius
$r$.
We define correlation integral $C_2$ as,
\begin{equation}
C_2(r)=\frac{1}{N}\sum\limits_{n=0}^{N}n P(n;r,N)
\label{eq:7}
\end{equation}
where $P(n;r,N)$ is the normalised probability of getting $n$ out of $N$
points as neighbours inside a radius $r$ of any of the points.
For a homogeneous distribution of points, the probability for any point to
fall within the neighbourhood is proportional to the ratio of the volume of
the sphere to the total volume of the sample.
In such a case $C_2$ reduces to the product of the volume of a sphere of
radius $r$ and the total number of particles, divided by the total volume.
As the total number and total volume are fixed quantities, $C_2$ for a
homogeneous distribution of points scales as $r^{D}$ at sufficiently large
scales.

The  power law scaling of correlation integral i.e. $C_2(r) \propto r^{D_2}$
defines the correlation dimension $D_2$ of the distribution.
\begin{equation}
D_2(r)=\frac{\partial \log C_2(r)}{\partial \log r}
\label{eq:78}
\end{equation}
Depending on the scaling of $C_2$, the value of correlation dimension $D_2$
can vary with scale $r$.
For the special case of a homogeneous distribution, we see that $D_2(r) =
D$ at sufficiently large scales and this matches the intuitive expectation
that the correlation dimension of a homogeneous distribution of points should
equal the mathematically defined fractal dimension for infinite number of
points.

We see that the correlation integral is defined in terms of
probability of finding $n$ point out of a distribution of $N$ points
within a distance $r$. This makes it a measure of one of the moments
of the distribution. We need all the moments of the distribution to
completely characterise the system statistically. The multi fractal
analysis used here does this with the generalised dimension $D_q$,
the Minkowski-Bouligand dimension, which is defined for an arbitrary
$q$ and typically computed for a range of values. It is different
from Renyi dimension only in the aspect that in this case the
spheres of radius $r$ have been centred at the point belonging to
the fractal whereas in Renyi dimension the sphere need not be
centred on the particle in the distribution (See section on Generalised
dimensions in \citet{1995PhR...251...1} for a discussion of the two types of
generalised dimensions.). 
The definition of generalised dimension $D_q$ is a generalisation of
the correlation dimension $D_2$. The correlation integral can be
generalised to define $C_q(r)$ as
\begin{equation}
C_q(r)=\frac{1}{NM}\sum_{i=1}^{M}n_i^{q-1}(r) =
\frac{1}{N}\sum\limits_{n=0}^{N}n^{q-1} P(n)
\label{eq:8}
\end{equation}
which is used to define the Minkowski-Bouligand dimension
\begin{equation}
D_q=\frac{1}{q-1}\frac{d\log{C_q(r)}}{d\log{r}}
\label{eq:9}
\end{equation}
The generalised dimension corresponds to the correlation dimension
for $q=2$.  
The values of
$C_q$ and $D_q$ can be related to a combination of correlation
functions for $q \geq 2$, with contribution from the two-point to
the $q$-point correlation functions for $C_q$. A multi fractal
structure, unlike a mono fractal can only be described by the full
spectrum of $D_q$. If the fractal in question is a mono-fractal then
we have $D_q = D_2$ for all $q$ and at all scales.

By construction, the positive values of $q$ give more weightage to regions
with a high number density whereas the negative values of $q$ give more
weightage to under dense regions.
Thus we may interpret $D_q$  for  $q \gg 0$ as characterising the scaling
behaviour of the galaxy distribution in the high density regions like clusters
whereas $q \ll 0$ characterises the scaling in voids.
In the situation where the galaxy distribution is homogeneous and isotropic
on large scales, we intuitively expect $D_q \simeq D=3$ independent of
the value of $q$ at the relevant scales.

\subsection{Homogeneous Distribution}

In our analysis, we first compute the expected values for $C_q$ and $D_q$ for
a homogeneous distribution in a finite volume.
The volume $V_{tot}$ over which the points are distributed is taken to be much
larger than volume of spheres ($V$).
The points are distributed randomly and we can use the Binomial distribution.
The probability of finding $n$ points in a sphere of volume $V$ centred on a
point, if $V_{tot}$ contains $N$ particles is:
\begin{equation}
P(n) =  \left({N-1\atop n-1}\right) p^{n-1} \left(1-p\right)^{N-n}
\label{eq:11}
\end{equation}
where $p$ is the probability that a given point (out of $N$) is located in a
randomly placed sphere.
The probability of finding only one particle in such a sphere is not equal to
$p$, in general.
If we place a sphere of volume $V$ inside a distribution which is contained in
volume $V_{tot}$ then $p = \frac{V}{V_{tot}}$.
We shall assume in our calculations that $p \ll 1$.
The above expression follows as with the cell centred on one point, this
point is already in the cell and we need to compute the probability of $n-1$
points out of $N-1$ being in the cell.
For comparison, the probability of finding $n$ particles in a randomly placed
sphere of volume $V$ is:
\begin{equation}
P(n) =  \left({N\atop n}\right) p^n \left(1-p\right)^{N-n}
\end{equation}
The average number of points in a randomly placed sphere is $\bar{N} = Np$ and
we assume that this is much larger than unity.
Thus we work in the situation where $1 \ll Np \ll N$.
Moments of the distribution for cells centred at points can be related to
moments for randomly placed cells.
\begin{eqnarray}
\langle \mathcal{N}^m \rangle_p &=& \sum n^m \left({N-1\atop
    n-1}\right) p^{n-1} \left(1-p\right)^{N-n} \nonumber \\
&=& \sum \left(n - 1 + 1\right)^m \left({N-1\atop n-1}\right) p^{n-1}
\left(1-p\right)^{N-n} \nonumber \\
&\simeq & \sum \left(n - 1\right)^m  \left({N-1\atop n-1}\right)
  p^{n-1} \left(1-p\right)^{N-n}  \nonumber \\
&& + \sum m \left(n-1\right)^{m-1} \left({N-1\atop n-1}\right) p^{n-1}
\left(1-p\right)^{N-n}  \nonumber \\
&\simeq& \langle \mathcal{N}^m\rangle + m\langle \mathcal{N}^{m-1} \rangle
\end{eqnarray}
The subscript $p$ on the angle brackets denotes that the average is for cells
centred on points within the distribution.
A specific application of the above expression is to compute the average
number of points in a spherical cell.
The average number of points in a sphere centred at a point is $ 1 +
(N-1)p \simeq Np + 1 = \bar{N} + 1$.
The difference between the two expressions arises due to fluctuations that are
present in an uncorrelated distribution of points.

The generalised correlation integral can now be expressed in terms of the
moments of this probability distribution.
In the limit $1 \ll Np \ll N$ we can write down a
leading order expression for the generalised correlation integral for $q > 1$
as:
\begin{equation}
NC_q(r) \simeq {\bar{N}}^{q-1} + \frac{(q-1)(q-2)}{2} {\bar{N}}^{q-2} +
(q-1){\bar{N}}^{q-2} + \cdots
\end{equation}
Here we have ignored terms that are
of lower order in $\bar{N}$ and terms of the same order in $\bar{N}$ with
powers of $p$ multiplying it.
(See Appendix A for a detailed discussion on how we arrived at this
expression.)
The Minkowski-Bouligand dimension corresponding to this is:
\begin{equation}
D_q(r) = D - \frac{\left( q - 2 \right)}{2} \frac{D }{\bar{N}} - \frac{D
}{\bar{N}}
\label{eqn:homogen}
\end{equation}
to the same order.
The last two terms in the intermediate expression for $D_q(r)$ have a
different origin: the first of the two terms arises due to fluctuations
present in a random distribution and the second term arises due to the cells
being centred at points within the distribution and this leads to weak
clustering.
A few points of significance are:
\begin{itemize}
\item
We do not expect $D_q(r)$ to coincide with the $D$ even if the
distribution of points is homogeneous.
Thus the benchmark for a sample of points is not $D$ but $D_q(r)$ given above,
and if the Minkowski-Bouligand dimension for a distribution of points
coincides with $D_q(r)$ then it may be considered as a homogeneous
distribution of points.
\item
The correction due to a finite size sample always leads to a smaller value for
$D_q(r)$ than the $D$.
\item
The correction is small if $\bar{N} \gg 1$, as expected.
The correction arises primarily due to discreteness and has been discussed by
\citet{1995PhR...251...1}.
The major advantage of our approach is that we are able to derive an
expression for the correction.
\end{itemize}

\subsection{Weakly Clustered Distribution}

We now consider weakly clustered distributions of points.
In this case the counts, for spheres whose centres are randomly placed and for
those centres are placed on the points in the distribution differ by a
significant amount.
Also there is no simple way of relating the two and hence we cannot use the
approach we followed in the previous subsection for estimating the generalised
correlation integral.

In order to make further progress, we note that we can always define an
average density for a distribution of a finite number of points in a finite
volume.
This allows us to go a step further and also define $n-$point correlation
functions.
It is well known that this can be used to relate the generalised correlation
integral with $n-$point correlation functions, e.g., see
\citet{1995PhR...251...1}.
We shall show below that it is possible to simplify this relation considerably
in the limit of weak clustering.
We can show that the correlation integral may be written as
follows (see Appendix B for details).
\begin{eqnarray}
NC_q(r) &\simeq& {\bar{N}}^{q-1}
\left( 1 +
  \frac{\left(q-1\right)\left(q-2\right)}{2{\bar{N}}} +
  \frac{q(q-1)}{2}\bar{\xi}\right.          \nonumber \\
&&+ \left. \mathcal{O}\left({\bar\xi}^2\right) +
\mathcal{O}\left(\frac{\bar\xi}{\bar{N}}\right)
+ \mathcal{O}\left(\frac{1}{{\bar{N}}^2}\right)\right)
\end{eqnarray}
Here we have used the assumption that $|\bar\xi| \ll 1$ and that higher powers
of $\bar\xi$ as well as higher order correlation functions can be ignored when
compared to terms of order $\bar\xi$ and $1/{\bar{N}}$.
This assumption is over and above the limit $1 \ll Np \ll N$.
The first two terms on the right hand side of this equation are same as the
first two terms in the expression for $C_q$ that we derived for a homogeneous
distribution of points.
The third term encapsulates the contribution of clustering.
This differs from the last term in the corresponding expression for a
homogeneous distribution as in that case the ``clustering'' is only due to
cells being centred at points whereas in this case the locations of every
pair of points has a weak correlation.
It is worth noting that the highest order term of order
$\mathcal{O}\left({\bar\xi}^2\right)$ has a factor
$\mathcal{O}\left(q^3\right)$ and hence can become important for sufficiently
large $q$.
{\it This may be quantified by stating that $q\bar\xi \ll 1$ is the more
  relevant small parameter for this expansion.}

The Minkowski-Bouligand dimension for such a system can now be expressed in
the form
\begin{eqnarray}
 D_q(r) &\simeq& D -
\frac{D \left( q - 2 \right)}{2\bar{N}} +
\frac{q}{2} \frac{\partial{\bar\xi}}{\partial{\log{r}}}   \nonumber \\
&=& D  -
\frac{D \left( q - 2 \right)}{2\bar{N}} -
\frac{Dq}{2} \left(\bar\xi(r) - \xi(r) \right) \nonumber  \\
&=& D - \left(\Delta{D_q}\right)_{\bar{N}} -
  \left(\Delta{D_q}\right)_{clus}
\label{eqn:clus}
\end{eqnarray}
It is interesting to see that the departure of $D_q$ from $D$ due to a
finite sample and weak clustering is given by distinct terms at the leading
order.
This expression allows us to compute $D_q$ for a distribution of points if the
number density and $\bar\xi$ are known.

Recall that $D$ is the mathematically defined fractal dimension for an
infinite set of points with a homogeneous distribution.
We have already noted some aspects of the correction due to a finite number of
points in the previous section, here we would like to highlight aspects of
corrections due to clustering.
\begin{itemize}
\item
For hierarchical clustering, both terms have the same sign and lead to a
smaller value for $D_q$ as compared to $D$.
\item
Unless the correlation function has a feature at some scale, smaller
correlation corresponds to a smaller correction to the Minkowski-Bouligand
dimension.
The expression given above quantifies this intuitive expectation.
\item
Note that for $q=2$, the expression given here is exact.  For this case, the
contribution of clustering has also been discussed by
\citet{1998MNRAS.298.1212M}.
\item
If the correlation function has a feature then it is possible to have a small
correction term $\left(\Delta{D_q}\right)_{clus}$ for a relatively large
$\xi$.
The relation between $\xi$ and $\left(\Delta{D_q}\right)_{clus}$ is not longer
one to one.
\end{itemize}

\begin{figure}
\includegraphics[width=3.2truein]{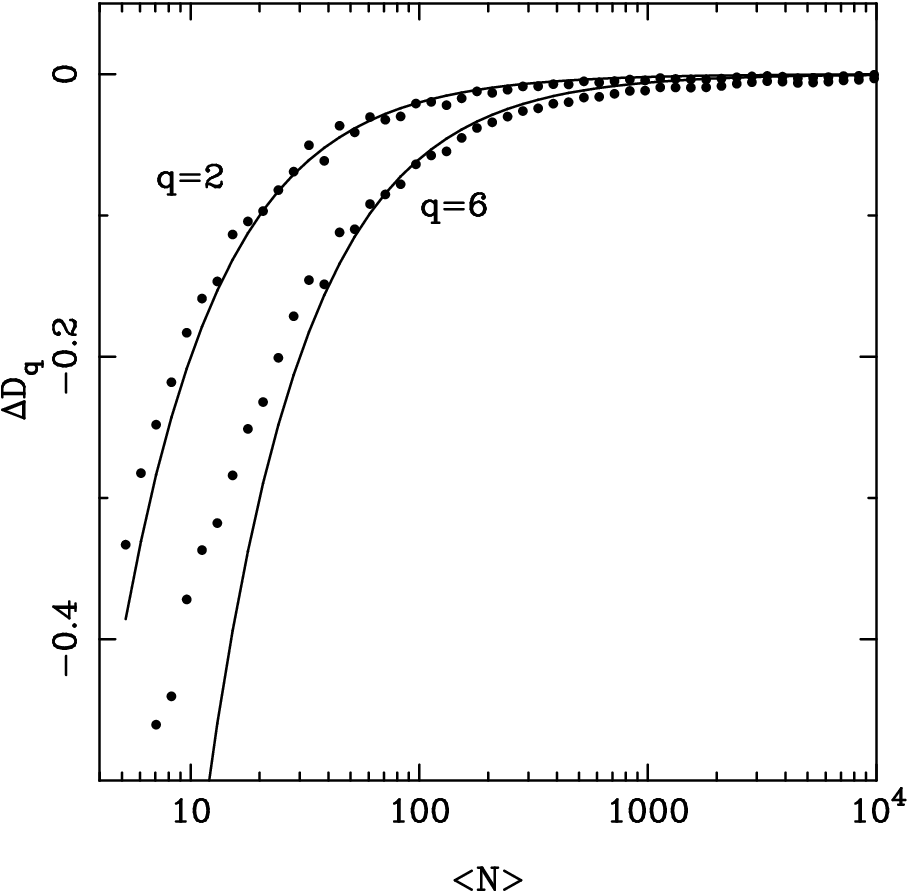}
\caption{Our model is compared with the observed Fractal dimensions for a
  random distribution of points in the special case of the multinomial model.
  $\Delta D_q$ is shown as a function of $\langle N \rangle \equiv \bar{N}$
  for $q=2$ and $6$ for this distribution.  $\Delta D_q$ measured from a
  realisation are plotted as points, and our model is shown as a curve.}
\end{figure}
\begin{figure}
\includegraphics[width=3.2truein]{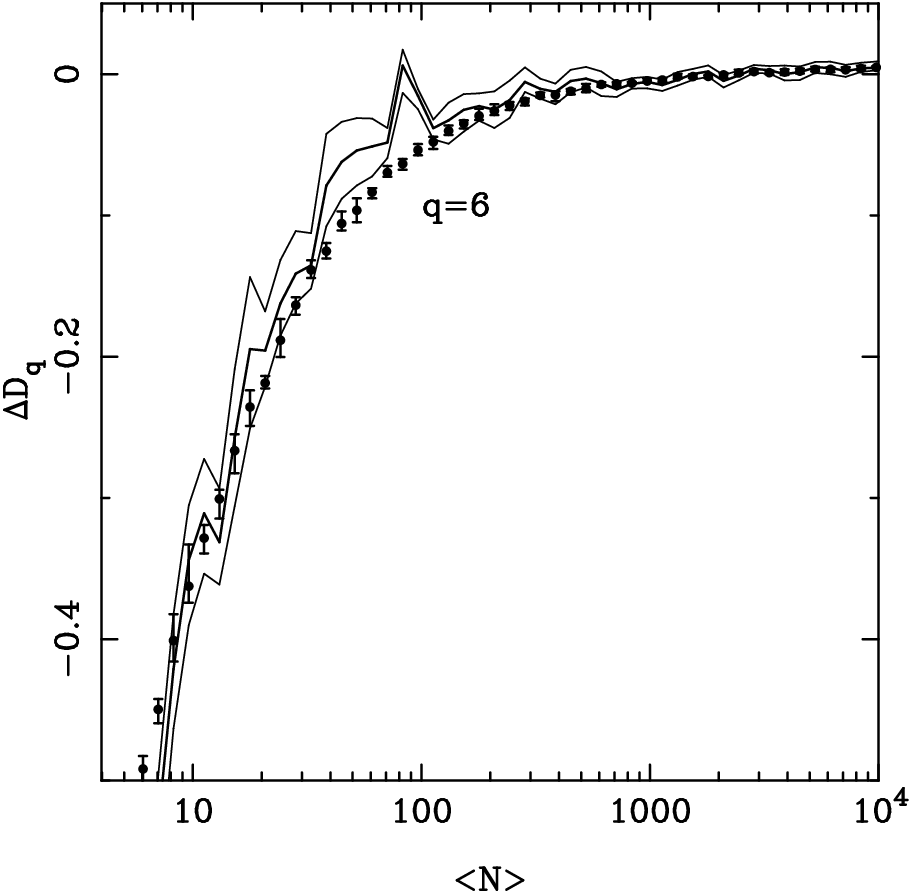}
\caption{Our model is compared with the observed Fractal dimensions for a
multinomial fractal with $f_i = {0.23,0.27,0.25,0.25}$.  $\Delta D_q \equiv
D_q - D_{q\, exp} $ is shown as a function of $\langle N \rangle$ for $q=6$.
We have plotted $\Delta D_q$ measured in the five realisations as points with
error bars.  The error bars mark the extreme values of $\Delta D_q$ seen in
these realisations whereas the central point marks the average value.
Predictions of our model based on correlation function measured in these
realisations is shown as a thick line.  This line corresponds to the average
value of $\xi$ and $\bar\xi$ measured in simulations, and thin lines mark the
predictions of our model based on extreme values seen in these simulations.}
\end{figure}
\begin{figure}
\includegraphics[width=3.2truein]{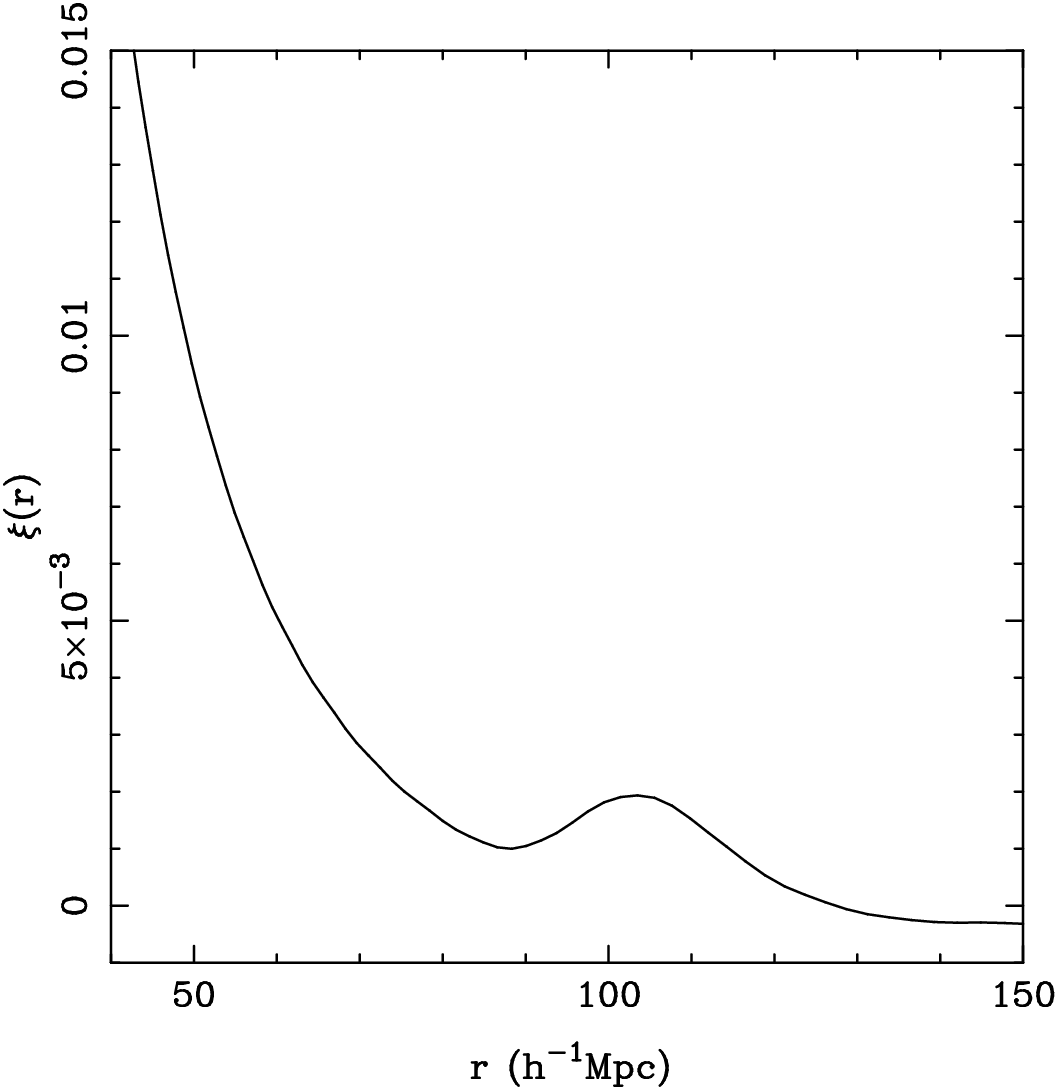}
\caption{The linearly extrapolated two point correlation
  function is shown as a function of scale for the best fit model for WMAP-3
  (see text for details).  This has been used, for calculation of $(\Delta
  D_q)_{clus}$.}
\end{figure}
\begin{figure}
\includegraphics[width=3.2truein]{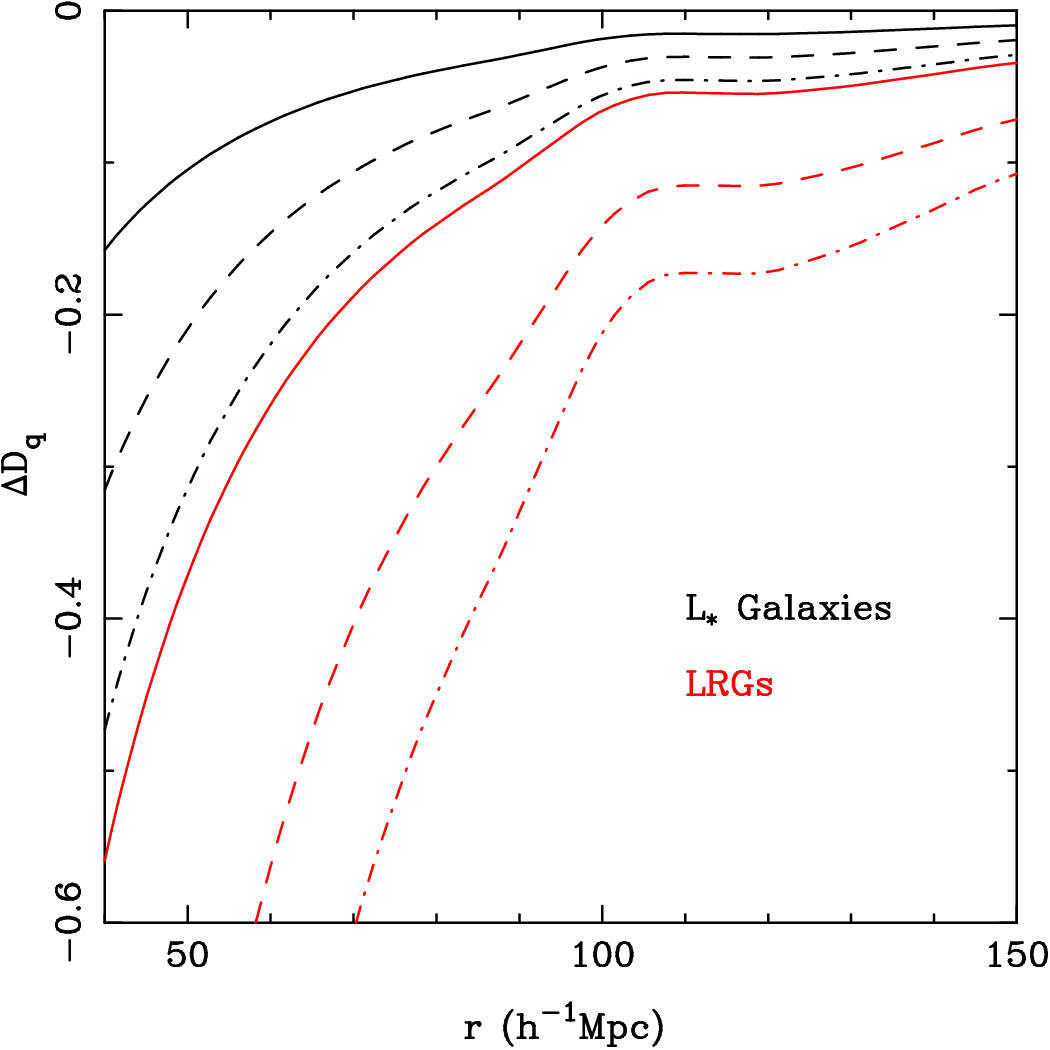}
\caption{Estimated deviation of the Minkowski-Bouligand dimension from the
  physical dimension is shown here for two types of populations.  In black we
  have plotted $\Delta D_q$ for an unbiased sample of points, distributed in
  redshift space with the real space correlation function as shown in
  Figure~1.  The solid curve shows $\Delta D_2$, whereas $\Delta D_4$ and
  $\Delta D_6$ are shown with a dashed curve and a dot-dashed curve
  respectively.  Curves in red correspond to an LRG like population with a
  number density of $5 \times 10^{-5}$~h$^{-3}$Mpc$^3$ and a linear bias of
  $2$.}
\end{figure}
\begin{figure}
\includegraphics[width=3.2truein]{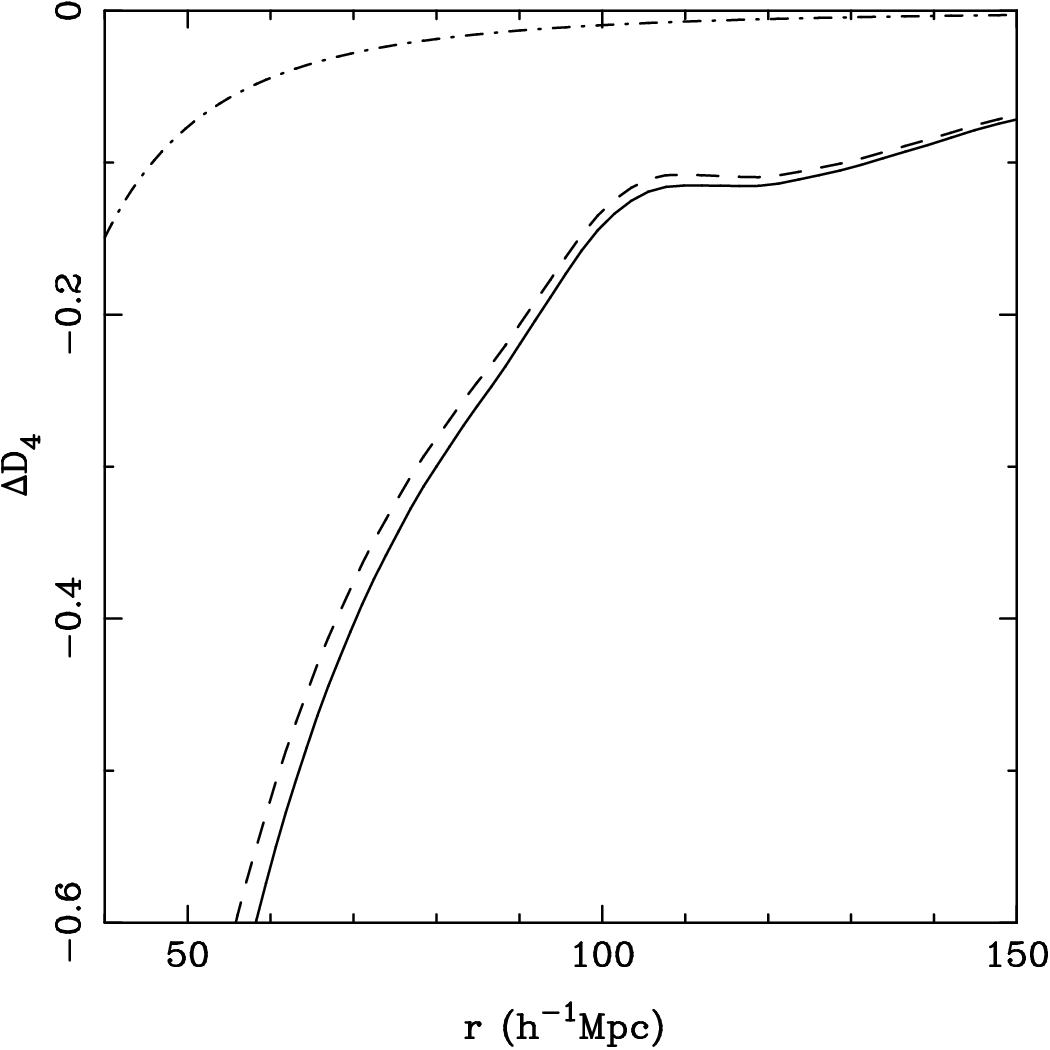}
\caption{This figure shows the components of $\Delta D_q$ for an LRG like
  population of galaxies for $q=4$.  This value of $q$ was chosen as the
  contribution of a finite number of galaxies does not vanish in this
  case. The solid line shows $\Delta D_4$, the dashed line shows the
  contribution of clustering to $\Delta D_q$ and the dot-dashed line is the
  correction due to a finite number of galaxies.  Clearly, the correction due
  to clustering is the dominant reason for departure of $D_q$ from $D$.}
\end{figure}

\subsection{Multifractal Multinomial Distribution}

We have applied our method to the multinomial multi fractal model
discussed in literature (See e.g. \citet{2002sgd..book.....M}).
The set of points for this model can be generated by starting with a
square and dividing it into four parts.
We assign a probability $\{f_i\}$ to each of these sub-squares
$\left(\sum\limits_{i=1}^{4}f_i=1\right)$.
This construction can be continued iteratively by dividing each smaller square
further and assigning probability by multiplying the corresponding number
${f_i}$ by all its ancestors.
We performed this construction to $L=8$ levels, thus getting a $256^2$ lattice
with the measure associated with each pixel.
For such models the we have an analytical expression for the generalised
dimension as
\begin{equation}
D_q=\frac{1}{q-1}\log_2\left(\sum\limits_{i=1,{f_i}\neq 0}^{4} {f_i}^q \right)
\end{equation}
This expression can be used to check whether our model for finite number and
correlation work correctly or not.

We have calculated the generalised dimension for this model taking three
different combination of ${f_i}$.
In one of the cases all four ${f_i}$'s are $0.25$ so that the distribution is
homogeneous.
In this case the expected $D_q = 2$ for all $q$ using the above expression.
Our model in this case gives a scale dependent correction to this due to a
finite number of particles.
Figure~1 shows $\Delta D_q$ as a function of $\langle N \rangle \equiv
\bar{N}$ for $q=2$ and $6$ for this distribution.
$\Delta D_q$ measured from a realisation are plotted as points, and our model
is shown as a curve.
It is clear that for $\bar{N} \leq 10^3$, there is a visible deviation of
$D_q$ from the expected value and that our model correctly estimates this
deviation.

In other case we present here, the ${f_i}$'s are close to $0.25$ but not
exactly equal to $0.25$, thus giving us a slightly clustered distribution.
We use $f_i = {0.23,0.27,0.25,0.25}$ and we generated give realisations on
this fractal.
In this case, the expected $D_q = 1.986$ for $q=6$.
As this differs from $D=2$, the difference in our model must come from
clustering presented in this fractal.
We generated five realisations of this fractal.
Figure~2 shows $\Delta D_q \equiv D_q - D_{q\, exp} $ as a function of
$\langle N \rangle$ for $q=6$, where $D_{q\, exp} $ follows from Eqn.(18).
We have plotted $\Delta D_q$ measured in the five realisations as points with
error bars.
The error bars mark the extreme values of $\Delta D_q$ seen in these
realisations whereas the central point marks the average value.
Predictions of our model (Eqn.(17)) based on correlation function measured in
these realisations is shown as a thick line.
This line corresponds to the average value of $\xi$ and $\bar\xi$ measured in
simulations, and thin lines mark the predictions of our model based on extreme
values seen in these simulations.
At $\langle N \rangle \ll 100$, where the effect of a finite number is
dominant, our model matches the measured $\Delta D_q$ very well.
At $\langle N \rangle \gg 100$ where the effect of clustering is dominant we
again find a good match between the model and measured values.
It is significant that at very large $\langle N \rangle$, we model the
deviation of $D_q$ from $D=2$ correctly.
However, there appears to be a mismatch in the transition region around
$\langle N \rangle \simeq 100$.
On inspection, we find that $\xi - \bar\xi$ has an oscillatory behaviour up to
this scale and the discrepancy corresponds to the last oscillation.
At the scale of maximum discrepancy, $\xi - \bar\xi \simeq 0.05$ and perhaps
we cannot ignore values of this order.

In summary we can say that our model works very well for the multinomial model
and we find that the correction due to clustering as well as a finite number
of points matches with the observed behaviour of $D_q$.

\section{Discussion}

The expressions derived in the previous section have a rich structure and we
illustrate some of the features here.
We would also like to discuss the application to the concordance model here.
The two point correlation function for the model that fits best the WMAP-3
data \citep{2007ApJS..170..377S} is shown in Figure~3.
We have used the flat $\Lambda$CDM model with a power law initial power
spectrum that best fits the WMAP-3 data here.
Parameters of the model used here are: $H_0=73$~km/s/Mpc, $\Omega_b h^2 =
0.0223$, $\Omega_c h^2 = 0.105$, $n_s=0.96$ and $\tau=0.088$.
For this model, $\sigma_8=0.76$.
The two point correlation has been shown at large scales where the clustering
can be assumed to be weak.
The most prominent feature here is the peak near $100$ Mpc.
This peak is caused by baryon acoustic oscillations (BAO) prior to
decoupling; see, e.g., \citet{1998ApJ...496..605E}.
Apart from this peak, the two point correlation function declines from small
scales towards larger scales at length scales shown here.

All observations of galaxies are carried out in redshift space.
Therefore we must use the correlation function in redshift space.
At large scales, redshift space distortions caused by infall lead to an
enhancement of the two point correlation function.
The enhancement is mainly along the line of sight but the angle averaged
two point correlation function is also amplified by some amount
\citep{1987MNRAS.227....1K}.

Further, we must also take into account the {\sl bias} in the distribution of
galaxies while using the correlation function shown in Figure~3.
This has been discussed by many authors
\citep{1984ApJ...284L...9K,1986ApJ...304...15B,1994ApJ...431..477B,1996ApJ...461L..65F,1996MNRAS.282..347M,1998MNRAS.297..251B,1998MNRAS.299..417B,1999ApJ...520...24D}.
At large scales, we may assume that the linear bias factor $b$ is sufficient
for describing the redshift space distortions and clustering.

Lastly, we should mention that we are working with the linearly extrapolated
correlation function at these scales even though there is some evidence that
perturbative effects lead to a slight shift in the location of the peak in
$\xi$ (For example, see \citet{2007astro.ph..3620S}).
The only change caused by such a shift in the location of the peak is to in
turn shift the scale where there appears to be a transition from large values
of $\Delta D_q$ towards small and constant values.
As the shift does not alter our key conclusions, we will ignore such effects
in the following discussion.

We plot the expected departure of $D_q$ from $D$ for an unbiased sample
of galaxies in Figure~4.
We assumed that typical (L*) galaxies have an average number density of
$0.02$~h$^3$Mpc$^{-3}$ and a bias factor of unity.
$\Delta D_q$ for such a population is shown as a function of scale by black
curves for $q=2$, $4$ and $6$.
Red curves show the same quantity for a sample of galaxies similar to Luminous
Red Galaxies (LRGs).
We used a bias factor $b=2$ and a number density of $5 \times
10^{-5}$~h$^3$Mpc$^{-3}$ that is representative of such a population.
For example, see \citet{2007MNRAS.381.1053P}.
$\Delta D_q$ is negative at all scales shown here, as expected from the
expression (see Eqn.(\ref{eqn:clus})) for hierarchical clustering.
The behaviour of $\Delta D_q$ as a function of scale has two distinct
regimes on either side of $100$~h$^{-1}$Mpc.
The magnitude of $\Delta D_q$ increases rapidly as we go from
$100$~h$^{-1}$Mpc towards smaller scales.
At scales larger than $100$~h$^{-1}$Mpc, $\Delta D_q$ either stays constant or
decreases at a very slow rate.
The behaviour of $\Delta D_q$ around $100$~h$^{-1}$Mpc is dictated largely by
the BAO peak in $\xi$ at this scale.
Although there is no peak in $\bar\xi$, $\partial \bar\xi / \partial\log r =
-0.5 D (\bar\xi(r) - \xi(r))$ has a minima and a maxima near the scale
of the peak in $\xi(r)$.
This results in a corresponding minima and maxima for $\Delta D_q$ as the
contribution of a finite number of galaxies is subdominant at such large
scales.
We illustrate this in Figure~5 where $\Delta D_4$ is plotted for an LRG like
sample, and the two contributions (from a finite sample and weak clustering)
are also shown.

If $\xi$ has a power law form then there are no extrema for $\partial
\bar\xi / \partial\log r$ and the magnitude of both $\xi$ and $\Delta D_q$
becomes progressively smaller as we get to larger scales.
There is a one to one relation between $\xi$ and $\Delta D_q$ for a given
model of this type.
However, a feature like the peak introduced by BAO leads to the non-trivial
behaviour illustrated in Figure~4.
Here we find that $D_q$ can be smaller at scales with a larger $\xi$.
For example, the scale with the local maxima of $\xi$ is very close to the
scale with the local minima of $D_q$.
The intuitive correspondence of a small $\xi$ implying a smaller deviation of
$D_q$ from $D$ does not apply in this case.

The difference between the unbiased galaxy population, and an LRG like sample
is stark.
The LRG like sample has a Minkowski-Bouligand dimension that differs from
$D=3$ by a significant amount.
The main reason for this difference is the high bias factor associated with
the LRG population, although a smaller number density also makes some
difference.
Different clustering properties for different types of galaxies imply that
these will have not have the same Minkowski-Bouligand dimension.
This has no impact on determination of the scale of homogeneity for the
universe, where we must use unbiased tracers.

The calculations presented in the previous section allow us to estimate the
offset of the Minkowski-Bouligand dimension from the physical dimension due to
weak clustering and a finite sample.
This has to be accompanied by a calculation of the dispersion in the expected
values \citep{1999MNRAS.310..428S,2000MNRAS.313..711C}.
The natural estimate for the scale of homogeneity is the scale where the
offset of the Minkowski-Bouligand dimension from the physical dimension
becomes smaller than the dispersion in a sufficiently large survey.
Given that the offset is dominated by the effect of clustering, we have
$\Delta D_q \simeq 0.5 D q (\xi - \bar\xi) \sim q (\xi - \bar\xi)$.
The offset scales with $q$.
Further, it is apparent that the dispersion in $\Delta D_q$ must also scale
with $q$.
This implies that the requirement of dispersion being greater than the offset
leads to the same scale for all $q$.
This is a very satisfying feature of this approach in that the scale of
homogeneity does not depend on the choice of $q$ as long as the effect of a
finite number of points is subdominant.

Alternatively, we may argue that the scale of homogeneity should be identified
with the scale above which the variation of $\Delta D_q$ is very small.
While this is an acceptable prescription for typical galaxies where $\Delta
D_q \leq 0.06$ at scales above $100$~h$^{-1}$Mpc, it does not appear
reasonable for an LRG like population.
The scale of homogeneity for the latter population is clearly much larger than
$100$~h$^{-1}$Mpc.

\section{Conclusions}

We have studied the problem of the expected value of the
Minkowski-Bouligand dimension for a finite distribution of points.
For this purpose, we have studied a homogeneous distribution as well as a
weakly clustered distribution.
In our study, $q/\bar{N}$ and $q\bar\xi$ are taken to be the
small parameters and the deviation of $D_q$ from $D$ is estimated in terms of
these quantities.
In both cases we find that the expected values of the Minkowski-Bouligand
dimension $D_q$ are different from  $D$ for the distribution of points.
For generic distributions, the value of $D_q$ is less than the dimension $D$.
We have derived an expression for $D_q$ in terms of the correlation function
and the number density in the limit of weak clustering.
{\it It is remarkable that $D_q < D$ even for homogeneous distributions. }

We find that $\Delta D_q = D_q - D$ is non-zero at all scales for
unbiased tracers of mass in the concordance model in cosmology.
For this model $\Delta D_q$ is a large negative number at small scales but it
rapidly approaches zero at larger scales.
$\Delta D_q$ is a very slowly varying function of scale above
$100$~h$^{-1}$Mpc and hence this may be tentatively identified as the scale of
homogeneity for this model.
A more quantitative approach requires us to estimate not only the systematic
offset $\Delta D_q$ but the dispersion in this quantity.
The scale of homogeneity can then be identified as the scale where the offset
is smaller than the expected dispersion.
We plan to undertake estimation of dispersion as the next step.
Verifying these results using simulated distributions of points is also on the
agenda.

Although we have used the example of galaxy clustering for illustrating our
calculations, the results as given in Eqn.(\ref{eqn:homogen}) and
Eqn.(\ref{eqn:clus}) are completely general and apply to any distribution of
points with weak departures from homogeneity.
A detailed derivation of the relations presented here, with verification using
mock distributions of points will be presented in a separate publication,
where we also expect to highlight other applications.


\section*{Acknowledgements}

The authors would like to thank Prof. K. Subramanian for useful comments.
JY is supported by a fellowship of the Council of Scientific and Industrial
Research (CSIR), INDIA.  
TRS thanks Department of Science and Technology, INDIA for financial
assistance.
JY and TRS acknowledge the facilities at the IUCAA Reference Centre at Delhi
University.
Computational work for this study was carried out at the cluster computing
facility in the Harish-Chandra Research Institute (http://cluster.hri.res.in).
We would like to thank the anonymous referee for useful comments.


\appendix

\section{Homogeneous Distribution}

The probability for a point to be found in a sphere of volume $V$ enclosed
within a total volume of $V_{tot}$ is $p=V/V_{tot}$.
In an uncorrelated distribution of points, the probability for all the points
are independent of one another and hence the probability of $n$ out of $N$
points falling in a sphere of volume $V$ is:
\begin{equation}
  P(n,N) = \left({N\atop n}\right)p^n (1-p)^{N-n}
\end{equation}
The distribution function determined by the probability function $P(n,N)$ is
called a Binomial distribution.
As discussed in the text, the probability distribution for occupation number
of cells centered at points is different but moments of that distribution can
be related to the moments of the probability distribution given above at the
required level of accuracy.
We require that the description should be accurate to first order in
$1/\bar{N}$.

The moment generating function for the Binomial distribution is given by
\begin{equation}
G(t)=\sum_{n=0}^N e^{tn}\left({N\atop n}\right)p^n(1-p)^{N-n}  = (pe^t+1-p)^N
\end{equation}
The $m$th moment of the distribution can then be calculated by differentiating
$G$ with respect to $t$, doing this $m$ times and then setting $t$ to zero.
The $m$th derivative of $G(t)$, at $t=0$ can be written as:
\begin{equation}
G^{(m)}(t)\left|_{t=0}\right. = \sum_{l=1}^{m}H_{m,l}\frac{N!}{(N-l)!}p^l
\end{equation}
where $H$ satisfies the following recurrence relation
\begin{equation}
H_{m,l}=l H_{m-1,l} + H_{m-1,l-1}
\end{equation}
with $H_{1,1}=1$ and $H_{m,l}=0$ for $l>m$ and $l<1$.
It can be shown that this implies $H_{l,l}=1$ for all $l$.

The $m$th moment of the distribution is given by:
\begin{eqnarray}
\langle \mathcal{N}^m \rangle &=& G^{(m)}(t)\left|_{t=0} \right. \nonumber \\
 &=& \sum_{l=1}^{m} H_{m,l}\frac{N!}{(N-l)!}p^l \nonumber
\end{eqnarray}
On the face of it this expression has a large number of terms for $m \gg 1$
and is difficult to analyse.
But if we assume that $p \ll 1$ and $\bar{N} = Np \gg 1$ then we can rewrite
the expression in the following form:
\begin{eqnarray}
\langle \mathcal{N}^m \rangle &=& H_{m,m}\frac{N!}{(N-m)!}p^{m} \nonumber \\
 && +H_{m,m-1}\frac{N!}{(N-m+1)!}p^{m-1} + \cdots \nonumber \\
 &=& \bar{N}^{m} + \mathcal{O}(p\bar{N}^{m-1})  \nonumber \\
&& + H_{m,m-1}\bar{N}^{m-1} +  \mathcal{O}(p\bar{N}^{m-2}) +
\mathcal{O}(\bar{N}^{m-2}) \nonumber \\
&\simeq& \bar{N}^{m}+\frac{m\left(m-1\right)}{2}\bar{N}^{m-1}
\end{eqnarray}
Where we have retained terms up to $\mathcal{O}(\bar{N}^{m-1})$ and have
dropped all other terms.
We have also used the recurrence relation and find that $H_{m,m-1}=
m(m-1)/2$.

We can now write the correlation integral as:
\begin{eqnarray}
NC_q(r) &=&  \langle \mathcal{N}^{q-1} \rangle + (q-1)\langle \mathcal{N}^{q-2}
\rangle \nonumber \\
&\simeq& {\bar{N}}^{q-1} + \frac{(q-1)(q-2)}{2} {\bar{N}}^{q-2} \nonumber \\
&& + (q-1){\bar{N}}^{q-2} + \cdots
\label{eq:app1}
\end{eqnarray}

The Minkowski-Bouligand dimension is then given by
\begin{eqnarray}
D_q(r)&=&\frac{1}{q-1}\frac{\partial \log C_q(r)}{\partial \log r} \nonumber \\
 &\simeq& \frac{1}{q-1}\frac{\partial }{\partial \log r}
 \log\left[\bar{N}^{q-1}\left(1 +
     \frac{\left(q-1\right)\left(q-2\right)}{2\bar{N}}
   \right)\right. \nonumber \\
&& \left. + \frac{\left(q-1\right)}{\bar{N}} \right]
 \nonumber \\
&\simeq& D\left(1-\frac{(q-2)}{2\bar{N}} - \frac{1}{\bar{N}}\right)
\label{eq:app3}
\end{eqnarray}
where $D$ is the dimension of the space in which particles are
distributed.
In this calculation, we have again made use of the fact that $\bar{N} \gg 1$
and that it scales as the $D$th power of scale $r$ for a random
distribution.


\section{Weakly Clustered Distribution}

In this section we will derive the form of the correlation integral for a
weakly clustered distribution of points.
Consider a sphere of volume $V$ contained within the sample of volume
$V_{tot}$.
We follow the approach given in \S{36} of \citet{1980lssu.book.....P} for
estimating the correlation integral.
In order to estimate the correlation integral, we divide the sphere into
infinitesimal elements such that each element contains at most one point.
This is a useful construct as $n_i^m=n_i$ for all $m \geq 0$, where $n_i$ is
the occupancy of the $i$th infinitesimal volume element.
If the occupancy of the $i$th volume element is $n_i$ then we have:
\begin{equation}
\mathcal{N} = \sum\limits_i n_i
\end{equation}
Therefore the mean count is:
\begin{equation}
\left\langle \mathcal{N} \right\rangle = \left\langle \sum\limits_i n_i
\right\rangle = \bar{N}
\end{equation}
The $m$th moment is then:
\begin{equation}
\left\langle {\mathcal{N}}^m \right\rangle = \left\langle \left(\sum\limits_i
    n_i \right)^m
\right\rangle
\end{equation}
If the sphere is centred at a point in the distribution then the averages are
denoted as $<\mathcal{N}^m>_p$, this is what we are interested in for the
purpose of computing the correlation integral.
\begin{equation}
\left\langle \mathcal{N}^m \right\rangle_p = \left\langle \left(\sum\limits_i
    n_i  \right)^m \right\rangle_p
\end{equation}
Averaging the sum raised to a positive integral power will lead to averaging
of terms of type $n_i n_j$, $n_i,n_j n_k$, etc. and the expression for such
terms involves $n$-point correlation functions, $n$ being the number of terms
being multiplied.
With this insight, we can write
\begin{eqnarray}
\left\langle \mathcal{N}^m \right\rangle_p  &=& \left\langle \sum n_1^m
\right\rangle_p + m \left\langle \sum n_1^{m-1} n_m
\right\rangle_p   + \cdots \nonumber \\
&& + \frac{m(m-1)}{2} \left\langle
  \sum n_1^2 n_3 \ldots n_m \right\rangle_p  \nonumber \\
&& + \left\langle
  \sum n_1 n_2 n_3 \ldots n_m \right\rangle_p \nonumber \\
&=& \left\langle \sum n_1
\right\rangle_p + m \left\langle \sum n_1 n_m
\right\rangle_p   + \cdots \nonumber \\
&& + \frac{m(m-1)}{2} \left\langle
  \sum n_1 n_3 \ldots n_m \right\rangle_p \nonumber \\
&& + \left\langle
  \sum n_1 n_2 n_3 \ldots n_m \right\rangle_p
\label{eq:app4}
\end{eqnarray}
Here terms in the expansion correspond to $i=j=k=\cdots$ for the first term,
only one of the indices differing from the rest for the second term and so
on.
The last term in this series is for all the $m$ indices different.
We have shifted the notation in order to write down the explicit form for
arbitrary $m$.

For a weakly clustered set of points with statistical isotropy and
homogeneity, we can safely assume that the magnitude of the two point
correlation function is small compared to unity, and higher order correlation
functions are even smaller.
Further, we continue to use the assumption that $\bar{N} \gg 1$ and hence we
need to retain only terms of the highest and the next highest order in this
parameter.
Thus we have two small parameters in the problem: $|\xi|$ and $1/\bar{N}$ and
our task is to compute the leading order terms in $\left\langle \mathcal{N}^m
\right\rangle_p$.
Here $\xi$ is the two point correlation function.

It can be shown that the leading order contribution comes from the last
term in the series in Eqn.(\ref{eq:app4}), and the next to leading order
contribution is from the last two terms.
We should note that these terms also contain several terms that are smaller
than the leading and next to leading order within them.
The foremost contribution comes from the uncorrelated component of the last
term, i.e., ${\bar{n}}^m \int dV_1 dV_2 \ldots dV_m = {\bar{N}}^m$.
The integral here is over $m$ independent volumes and $\bar{n}$ is the average
number density.
The next contribution comes from components of this term that include the
effect of pairwise correlations.
As there are $m$ distinct points, the number of distinct pairs is $m(m+1)/2$
and the term has the form:
\begin{equation}
{\bar{N}}^m \frac{m(m+1)}{2} \bar\xi(r)
\end{equation}
where $r$ is the radius of the sphere with volume $V$ and $\bar\xi$ is given
by:
\begin{equation}
\bar\xi(r) = \frac{3}{r^3} \int\limits_0^r x^2 \xi(x) dx ~~~~~~~ .
\end{equation}
It can be shown that all other components of the last term involve higher
order correlation functions, or higher powers of $\xi$.
Further, it can be shown that the contributions that contain only a single
power of $\xi$ from other terms in the series in Eqn.(\ref{eq:app4}) contain a
lower power of $\bar{N}$.
Lastly, it can be shown that the only other term that we need to take into
account comes from the penultimate term in the series in Eqn.(\ref{eq:app4}).
The uncorrelated component of this term is:
\begin{equation}
\frac{m(m-1)}{2} {\bar{N}}^{m-1}
\end{equation}
Thus we have for the $m$th moment of the counts of neighbours:
\begin{eqnarray}
\left\langle \mathcal{N}^m \right\rangle_p &=& \bar{N}^m + \frac{m(m+1)}{2}
\bar{N}^m \bar{\xi}+\frac{m(m-1)}{2} \bar{N}^{m-1}  \nonumber \\
&& + \bar{N}^m \left(\mathcal{O}\left({\bar\xi}^2\right) +
  \mathcal{O}\left(\frac{\bar\xi}{\bar{N}}\right)
+ \mathcal{O}\left(\frac{1}{{\bar{N}}^2}\right) \right) \nonumber \\
&\simeq&
\bar{N}^m\left(1+\frac{m(m+1)}{2}\bar{\xi}+\frac{m(m-1)}{2\bar{N}}\right)
\label{eq:appb2}
\end{eqnarray}
The largest term of order $\mathcal{O}\left({\bar\xi}^2\right)$ arises from
the contribution of correlated triangles in the last term of
Eqn.(\ref{eq:app4}).
The number of triangles scales as $m^3$ and hence can become important for
sufficiently large $m$.
This may be codified by stating that $m\bar\xi \ll 1$ is the more relevant
small parameter.

The correlation integral can be written as
\begin{equation}
NC_q(r) \simeq \bar{N}^{q-1}\left(1+\frac{q(q-1)}{2}\bar{\xi}
  +\frac{(q-1)(q-2)}{2\bar{N}}\right)
\label{eq:appb3}
\end{equation}
From this we can calculate the Minkowski-Bouligand dimension using equation
\ref{eq:app3} as
\begin{eqnarray}
 D_q(r) &=& \frac{1}{(q-1)} \frac{\partial\log C_q(r)}{\partial\log
   r}      \nonumber \\
&\simeq& D -
\frac{D \left( q - 2 \right)}{2\bar{N}} +
\frac{q}{2} \frac{\partial{\bar\xi}}{\partial{\log{r}}}   \nonumber \\
&=& D  -
\frac{D \left( q - 2 \right)}{2\bar{N}}   - \frac{D q}{2} \left(\bar\xi(r) - \xi(r) \right)
\end{eqnarray}
This is the required expression.

\label{lastpage}


\begin{thebibliography}{99}

\bibitem[\protect\citeauthoryear{Amendola \&
Palladino}{1999}]{1999ApJ...514L...1A} Amendola L., Palladino E., 1999,
ApJ, 514, L1

\bibitem[Bagla(1998a)]{1998MNRAS.297..251B} Bagla, J.~S.\ 1998a, \mnras, 297,
251

\bibitem[Bagla(1998b)]{1998MNRAS.299..417B} Bagla, J.~S.\ 1998b, \mnras, 299,
417

\bibitem[\protect\citeauthoryear{Bardeen et
al.}{1986}]{1986ApJ...304...15B} Bardeen J.~M., Bond J.~R., Kaiser N.,
Szalay A.~S., 1986, ApJ, 304, 15

\bibitem[\protect\citeauthoryear{Baryshev \&
Bukhmastova}{2004}]{2004AstL...30..444B} Baryshev Y.~V., Bukhmastova Y.~L.,
2004, AstL, 30, 444

\bibitem[Benoist et al.(1996)]{1996ApJ...472..452B} Benoist, C.,
Maurogordato, S., da Costa, L.~N., Cappi, A.,
\& Schaeffer, R.\ 1996, \apj, 472, 452

\bibitem[Bernardeau et al.(2002)]{2002PhR...367....1B} Bernardeau, F.,
Colombi, S., Gazta{\~n}aga, E., \& Scoccimarro, R.\ 2002, \physrep, 367, 1

\bibitem[\protect\citeauthoryear{Best}{2000}]{2000ApJ...541..519B} Best
J.~S., 2000, ApJ, 541, 519

\bibitem[\protect\citeauthoryear{Bharadwaj, Gupta, \&
Seshadri}{1999}]{1999A&A...351..405B} Bharadwaj S., Gupta A.~K., Seshadri
T.~R., 1999, A\&A, 351, 405

\bibitem[\protect\citeauthoryear{Borgani et
al.}{1993}]{1993PhRvE..47.3879B} Borgani S., Murante G., Provenzale A.,
Valdarnini R., 1993, PhRvE, 47, 3879

\bibitem[\protect\citeauthoryear{Borgani \&
Murante}{1994}]{1994PhRvE..49.4907B} Borgani S., Murante G., 1994, PhRvE,
49, 4907

\bibitem[\protect\citeauthoryear{Borgani}{1995}]{1995PhR...251...1} Borgani
  S.,  1995, \physrep,  251, 1

\bibitem[\protect\citeauthoryear{Brainerd \&
Villumsen}{1994}]{1994ApJ...431..477B} Brainerd T.~G., Villumsen J.~V.,
1994, ApJ, 431, 477

\bibitem[\protect\citeauthoryear{Cappi et al.}{1998}]{1998A&A...335..779C}
Cappi A., Benoist C., da Costa L.~N., Maurogordato S., 1998, A\&A, 335, 779

\bibitem[Cole et al.(2005)]{2005MNRAS.362..505C} Cole, S., et al.\ 2005,
\mnras, 362, 505

\bibitem[Cole et al.(2006)]{2006astro.ph.11178C} Cole, S., Sanchez, A.~G.,
\& Wilkins, S.\ 2006, ArXiv Astrophysics e-prints, arXiv:astro-ph/0611178

\bibitem[\protect\citeauthoryear{Coleman \&
Pietronero}{1992}]{1992PhR...213..311C} Coleman P.~H., Pietronero L., 1992,
PhR, 213, 311

\bibitem[Colless et al.(2001)]{2001MNRAS.328.1039C} Colless, M., et al.\
2001, \mnras, 328, 1039

\bibitem[Colombi et al.(2000a)]{2000PhRvL..85.5515C} Colombi, S., Pogosyan,
D., \& Souradeep, T.\ 2000a, Physical Review Letters, 85, 5515

\bibitem[\protect\citeauthoryear{Colombi et
al.}{2000b}]{2000MNRAS.313..711C} Colombi S., Szapudi I., Jenkins A.,
Colberg J., 2000b, MNRAS, 313, 711

\bibitem[\protect\citeauthoryear{Dekel \&
Lahav}{1999}]{1999ApJ...520...24D} Dekel A., Lahav O., 1999, ApJ, 520, 24


\bibitem[Dressler(1980)]{1980ApJ...236..351D} Dressler, A.\ 1980, \apj,
236, 351

\bibitem[Einstein(1917)]{1917SPAW.......142E} Einstein, A.\ 1917,
Sitzungsberichte der K{\"o}niglich Preu{\ss}ischen Akademie der
Wissenschaften (Berlin), Seite 142-152., 142

\bibitem[Eisenstein \& Hu(1998)]{1998ApJ...496..605E} Eisenstein, D.~J., \&
Hu, W.\ 1998, \apj, 496, 605

\bibitem[\protect\citeauthoryear{Fry}{1996}]{1996ApJ...461L..65F} Fry
J.~N., 1996, ApJ, 461, L65

\bibitem[\protect\citeauthoryear{Guzzo}{1997}]{1997NewA....2..517G} Guzzo
L., 1997, NewA, 2, 517

\bibitem[\protect\citeauthoryear{Hatton}{1999}]{1999MNRAS.310.1128H} Hatton
S., 1999, MNRAS, 310, 1128

\bibitem[\protect\citeauthoryear{Hawkins et
al.}{2003}]{2003MNRAS.346...78H} Hawkins E., et al., 2003, MNRAS, 346, 78

\bibitem[\protect\citeauthoryear{Hogg et al.}{2005}]{2005ApJ...624...54H}
Hogg D.~W., Eisenstein D.~J., Blanton M.~R., Bahcall N.~A., Brinkmann J.,
Gunn J.~E., Schneider D.~P., 2005, ApJ, 624, 54

\bibitem[\protect\citeauthoryear{Kaiser}{1984}]{1984ApJ...284L...9K} Kaiser
N., 1984, ApJ, 284, L9

\bibitem[\protect\citeauthoryear{Jones et al.}{2005}]{2005RvMP...76.1211J}
Jones B.~J., Mart{\'{\i}}nez V.~J., Saar E., Trimble V., 2005, RvMP, 76,
1211

\bibitem[Kaiser(1987)]{1987MNRAS.227....1K} Kaiser, N.\ 1987, \mnras, 227,
1

\bibitem[Kim et al.(2002)]{2002AJ....123...20K} Kim, R.~S.~J., et al.\
2002, \aj, 123, 20

\bibitem[Kulkarni et al.(2007)]{2007MNRAS.378.1196K} Kulkarni, G.~V.,
Nichol, R.~C., Sheth, R.~K., Seo, H.-J., Eisenstein, D.~J., \& Gray, A.\
2007, \mnras, 378, 1196

\bibitem[\protect\citeauthoryear{Kurokawa, Morikawa, \&
Mouri}{2001}]{2001A&A...370..358K} Kurokawa T., Morikawa M., Mouri H.,
2001, A\&A, 370, 358

\bibitem[\protect\citeauthoryear{Lahav}{2002}]{2002CQGra..19.3517L} Lahav
O., 2002, CQGra, 19, 3517

\bibitem[de Lapparent et al.(1986)]{1986ApJ...302L...1D} de Lapparent, V.,
Geller, M.~J., \& Huchra, J.~P.\ 1986, \apjl, 302, L1

\bibitem[de Lapparent \& Slezak(2007)]{2007A&A...472...29D} de Lapparent,
V., \& Slezak, E.\ 2007, \aap, 472, 29

\bibitem[Mandelbrot(1982)]{1982fgn..book.....M} Mandelbrot, B.~B.\ 1982,
The Fractal Geometry of Nature, San Francisco: Freeman, 1982,

\bibitem[\protect\citeauthoryear{Martinez}{1999}]{1999Sci...284..445M}
Martinez V.~J., 1999, Sci, 284, 445

\bibitem[\protect\citeauthoryear{Martinez et
al.}{1998}]{1998MNRAS.298.1212M} Martinez V.~J., Pons-Borderia M.-J.,
Moyeed R.~A., Graham M.~J., 1998, MNRAS, 298, 1212

\bibitem[Mart{\'{\i}}nez et al.(2001)]{2001ApJ...554L...5M}
Mart{\'{\i}}nez, V.~J., L{\'o}pez-Mart{\'{\i}}, B.,
\& Pons-Border{\'{\i}}a, M.-J.\ 2001, \apjl, 554, L5

\bibitem[Mart{\'{\i}}nez \& Saar(2002)]{2002sgd..book.....M} Mart{\'{\i}}nez,
  V.~J., \& Saar, E.\ 2002, Statistics of the Galaxy Distribution, Published
  by Chapman \& Hall/CRC, Boca Raton, ISBN:
1584880848,

\bibitem[\protect\citeauthoryear{Mo \& White}{1996}]{1996MNRAS.282..347M}
Mo H.~J., White S.~D.~M., 1996, MNRAS, 282, 347

\bibitem[Padmanabhan(2002)]{2002tagc.book.....P} Padmanabhan, T.\ 2002,
Theoretical Astrophysics, by T.~Padmanabhan, pp.~638.~ISBN
0521562422.~Cambridge, UK: Cambridge University Press, October 2002.,

\bibitem[\protect\citeauthoryear{Pan \& Coles}{2000}]{2000MNRAS.318L..51P}
Pan J., Coles P., 2000, MNRAS, 318, L51

\bibitem[Peacock(1999)]{1999coph.book.....P} Peacock, J.~A.\ 1999,
Cosmological Physics, by John A.~Peacock, pp.~704.~ISBN
052141072X.~Cambridge, UK: Cambridge University Press, January 1999.,

\bibitem[\protect\citeauthoryear{Peebles}{1980}]{1980lssu.book.....P}
Peebles P.~J.~E., 1980, {\sl Large Scale Structure in the Universe},
Princeton University Press, Princeton, USA

\bibitem[Percival et al.(2007a)]{2007MNRAS.381.1053P} Percival, W.~J., Cole,
S., Eisenstein, D.~J., Nichol, R.~C., Peacock, J.~A., Pope, A.~C., \&
Szalay, A.~S.\ 2007a, \mnras, 381, 1053

\bibitem[Percival et al.(2007b)]{2007ApJ...657..645P} Percival, W.~J., et
al.\ 2007b, \apj, 657, 645

\bibitem[\protect\citeauthoryear{ Pietronero}{1987}]{1987PhyA..144..257P}
  Pietronero L., \ 1987, Physica A, 144, 257

\bibitem[Ross et al.(2006)]{2006astro.ph.12400R} Ross, N.~P., et al.\ 2006,
ArXiv Astrophysics e-prints, arXiv:astro-ph/0612400

\bibitem[Sanchez \& Cole(2007)]{2007arXiv0708.1517S} Sanchez, A.~G., \&
Cole, S.\ 2007, ArXiv e-prints, 708, arXiv:0708.1517

\bibitem[Shectman et al.(1996)]{1996ApJ...470..172S} Shectman, S.~A.,
Landy, S.~D., Oemler, A., Tucker, D.~L., Lin, H., Kirshner, R.~P., \&
Schechter, P.~L.\ 1996, \apj, 470, 172

\bibitem[Smith et al.(2007)]{2007astro.ph..3620S} Smith, R.~E.,
Scoccimarro, R., \& Sheth, R.~K.\ 2007, ArXiv Astrophysics e-prints,
arXiv:astro-ph/0703620

\bibitem[Spergel et al.(2007)]{2007ApJS..170..377S} Spergel, D.~N., et al.\
2007, \apjs, 170, 377

\bibitem[\protect\citeauthoryear{Sylos Labini \rm{et al.}
  }{1998}]{1998PhR...293...61S} Sylos Labini, F., Montuori, M.,  Pietronero,
  L.\ 1998, \physrep, 293, 61

\bibitem[Sylos Labini et
al.(2007)]{2007A&A...465...23S} Sylos Labini, F., Vasilyev, N.~L.,
\& Baryshev, Y.~V.\ 2007, \aap, 465, 23

\bibitem[\protect\citeauthoryear{Szapudi, Colombi, \&
Bernardeau}{1999}]{1999MNRAS.310..428S} Szapudi I., Colombi S., Bernardeau
F., 1999, MNRAS, 310, 428

\bibitem[van de Weygaert \& Schaap(2007)]{2007arXiv0708.1441V} van de
Weygaert, R., \& Schaap, W.\ 2007, ArXiv e-prints, 708, arXiv:0708.1441

\bibitem[Vasilyev et
al.(2006)]{2006A&A...447..431V} Vasilyev, N.~L., Baryshev, Y.~V., \&
Sylos Labini, F.\ 2006, \aap, 447, 431

\bibitem[\protect\citeauthoryear{Wu, Lahav, \&
Rees}{1999}]{1999Natur.397..225W} Wu K., Lahav O., Rees M., 1999, Nature,
397, 225

\bibitem[\protect\citeauthoryear{Yadav \rm{et
      al.}}{2005}]{2005MNRAS.364..601Y} Yadav J, Bharadwaj S., Pandey B,
  Seshadri T.R.,\ 2005, \mnras, 364, 601

\bibitem[York et al.(2000)]{2000AJ....120.1579Y} York, D.~G., et al.\ 2000,
\aj, 120, 1579

\bibitem[\protect\citeauthoryear{Zehavi \rm{et
      al.}}{2002}]{2002ApJ...571..172Z} Zehavi I. \rm{et al.}, 2002,
  \apj,571,172

\end{thebibliography}
\end{document}